# Long-period fiber grating as wavelength selective element in double-clad Yb-doped fiber-ring lasers

P. Peterka[a], J. Maria[b], B. Dussardier[b], R. Slavík[a,c] P. Honzátko[a], and V. Kubeček[d]

[a]*Institute of Photonics and Electronics, Academy of Sciences of the Czech Republic, Chaberská 57, 18251 Prague, Czech Republic*
[b]*Laboratoire de Physique de la Matière Condensée, CNRS - Université de Nice - Sophia Antipolis, 06108, Nice, France*
[c]*Now with Optoelectronic Research Centre, University of Southampton, Southampton, SO17 1BJ, United Kingdom*
[d]*Czech Technical University, Faculty of Nuclear Sciences and Physical Engineering, Břehová 7, 115 19 Prague, Czech Republic*
e-mail: peterka@ufe.cz

**Abstract:** Selection of operating wavelength of the Yb-doped fiber-ring lasers using long-period fiber gratings (LPFGs) is suggested. In the proposed method, customized LPFG that sustains high powers serves as a broad-band rejection filter. It modifies the net gain profile of the laser, enabling the peak gain to occur at a designed wavelength. Spectral range of oscillation between 1050-1110 nm was experimentally demonstrated. This range can be extended to both shorter and longer wavelengths with proper design of the LPFG and length of the Yb-doped fiber. The gratings were inscribed by $CO_2$ laser and the grating period down to 175 μm was achieved being, to our best knowledge, the shortest reported LPFG period using this technique.

## 1. Introduction

Cladding-pumped ytterbium-doped fiber lasers belong presently among the most attractive high-power laser sources for wavelengths around 1060 nm. While the most typical configuration of the fiber laser cavity is linear Fabry-Perot arrangement with fiber Bragg gratings (FBGs), for some applications the ring cavity is preferable as it offers higher stability when unidirectional operation is enforced with fiber optic isolator [Hideur00]. Indeed, Yb-doped fiber lasers are prone to the so-called sustained self-pulsing that in the case of highly Yb-doped fiber (YDF) may under certain circumstances lead to relatively stable self-Q-switched operation [Kir'yanov06]. Particular application we have in mind is testing of saturable absorbers based on $Cr^{4+}$-doped fiber for passively Q-switched (PQS) all-fiber laser. The high-power Q-switched sources in all-fiber configuration are of increasing interests [Jiang09, Kurkov09, Andres08]. While the $Cr^{4+}$-doped crystals are widely used as a Q-switching elements in bulk or micro-chip solid state lasers [Dong05], Q-switched operation due to $Cr^{4+}$-doped fiber was successfully demonstrated only in case of core-pumped low-power Nd-doped fiber laser [Tordella03]. In the PQS fiber laser with the cladding-pumped YDF a significant increase of the average output power can be expected. To distinguish between the PQS regime caused by the saturable absorber and the YDF itself, we have built more stable laser in ring configuration.

The possibility of selection of the laser wavelength in a wide spectral range is also important as the optimum operation wavelength of the saturable absorber does not necessarily correspond to the peak gain of the gain medium [Kurkov09]. However, selection of the laser output wavelength in ring resonators by the narrow-band reflective filters (e.g., FBGs) requires fiber-optic circulator [Chen06]. Transmission filters like thin film filters are prone to damage under high optical power. An alternative technology is based on long-period fiber gratings (LPFGs) – they have very high damage threshold and operate in transmission. The challenge in using LPFGs as wavelength selective element is in the fact that they typically form a stop-band filter rather than a pass-band filter. Up to date, LPFG were used mainly in the Fabry-Perot configuration: e.g., to suppress gain of the Nd-doped fiber Fabry-Perot laser at ~1060 nm, allowing lasing at ~1088 nm, defined by a FBG [Aslund07] and mechanically induced LPFG was used for subtle tuning of an Yb-doped fiber Fabry-Perot laser [Anzueto-Sanchez07]. Yan and co-workers used spectral comb filter based on two LPFGs inscribed slightly apart onto single mode fiber for building a triple wavelength switchable erbium doped fiber ring laser [Yan07]. But the spectral filtering was realized thanks to the comb-like spectral transmission function with spectral period of 2.6 nm of the Mach-Zehnder interferometer made of two LPFGs and not by the spectral transmission of the individual LPFGs. In fact, all three wavelengths coincided with the peak gain of the laser cavity. Here, we present an Yb-doped fiber-ring laser with wavelength selected in a wide spectral range by one or more LPFGs. By proper design of the LPFGs, their transmission modifies the net gain profile of the laser, enabling the peak gain to occur at a desired wavelength.

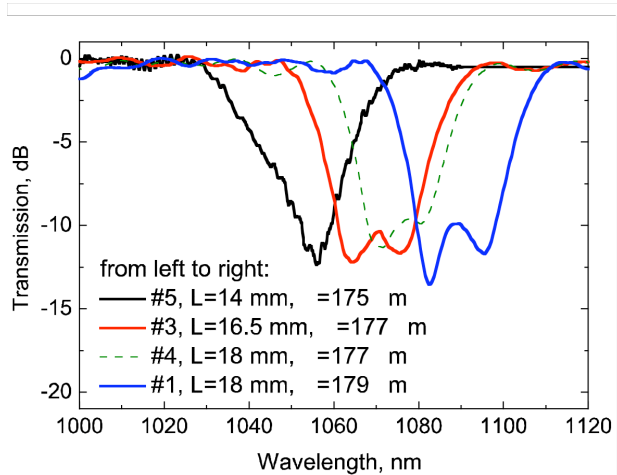
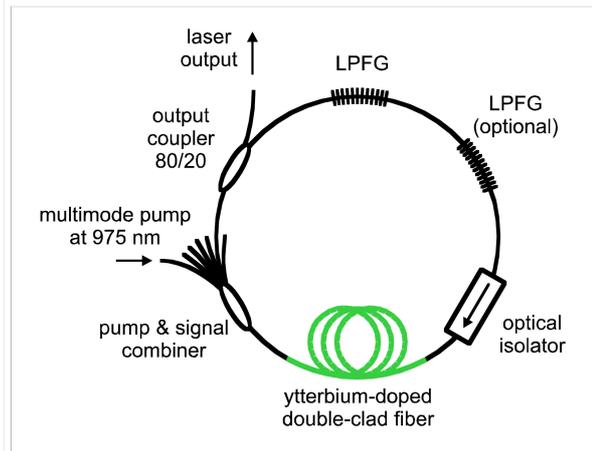

**Figure 1** Transmission of the developed LPFGs.   **Figure 2** Setup of the fiber laser.

## 2. Long-period fiber grating fabrication and characteristics

The LPFGs were inscribed onto the fiber by $CO_2$ laser [Grubsky06, Slavik09a, Slavik09b], which is a technique that allows for much higher temperature stability than, e.g., UV inscription. Consequently, very high power handling typical for fiber gratings (FBGs, LPFGs) is further increased. The LPFGs were prepared in a passive fiber (Fibercore PS1250/1500). Although the fiber is not single-mode at the laser signal wavelength, it can be assumed that only fundamental mode is excited as the fiber is only about half a meter long and it is spliced between two fibers that are single-mode in the spectral region of interest. The gratings have period of 175-179 µm (the shortest period reported with $CO_2$ laser technique so far) and length of 14-18 mm. Spectral characteristics of several LPFG samples are shown in Fig. 1.

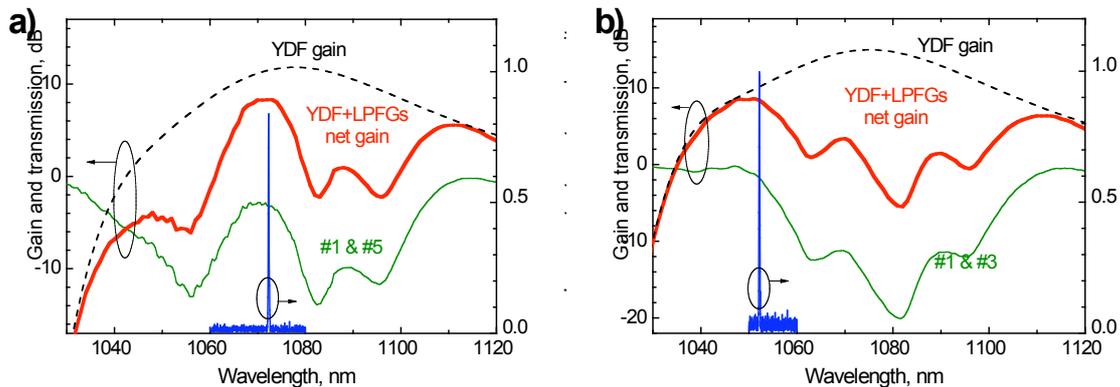

**Figure 3** Principle of selection of the laser wavelength by combination of appropriate spectral transmission of the LPFGs and YDF gain: (a) lasing occurs in a transmission window between two LPFGs or (b) on a side of attenuation band of the LPFGs.

## 3. Laser experiment and discussion

The laser setup is shown in Fig. 2. The gain medium was a 3 m-long double-clad YDF (Liekki-nLight Yb1200-6/125DC) with nominal 2.6 dB/m peak absorption in the inner cladding and 1200 dB/m in the core. The inner cladding has an octagonal shape with 125 µm average diameter and the core has the diameter of 6 µm and numerical aperture of 0.13. The outer cladding is made from low refractive index polymer. The optical isolator capable of optical power handling up to 5 W (cw regime) or 5 kW (pulsed regime) ensured unidirectional laser operation. The total insertion losses of the cavity are estimated to be 1.5 dB, mainly caused by the isolator (0.8 dB), pump and signal combiner (0.5 dB) and splice losses. The laser output was obtained from a fused optical fiber coupler 80:20. If not otherwise stated, the 80% branch of the coupler was used for the output. In the experiments, we used one pump laser diode pigtailed with multimode fiber of 105 µm core diameter, but up to six laser diodes can be used. Pump power of 4.4 W in maximum was coupled into the inner cladding of the YDF. The laser diode wavelength was tuned to the Yb peak absorption by temperature so that the absorbed pump power is maximal at the highest allowed diode current of 7 A. Without any wavelength selective component, the laser oscillated in the range of 1076-1084 nm, and we observed wavelength drift with time. In Fig. 3, we illustrate on two examples the principle of laser wavelength selection by insertion of one or more LPFGs. In Fig. 3a, the lasing wavelength of 1072.3 nm is set by the transmission window of the two concatenated LPFGs #1 and #5. The actual laser wavelength is given by the maximum net gain given as a product of the LPFGs transmission and YDF gain. The large-signal fiber gain in the Fig. 3 was estimated using numerical model of the fiber laser

[Peterka04, Peterka08]. In the example shown in Fig. 3b, the laser is forced to oscillate at 1053.2 nm, at the short wavelength edge of the transmission of LPFG #3. The LPFG #1 is inserted into the cavity to block the amplified spontaneous emission (ASE) and to prevent lasing at the long wavelength transmission edge of the LPFG #3. Indeed, if the LPFG #1 is not present, the laser wavelength was 1094 nm as can be seen in Fig. 4a.

In the case of lower inversion of excited level population, e.g., for longer YDF or lower output coupler ratio, the fiber gain is shifted significantly towards longer wavelengths. In order to investigate the regime of lower inversion of the active medium we interchanged the branches of the output coupler and the 20% branch was used as the output. In such a case, the combination of LPFGs #1 and #3 from the example in Fig. 3b lead to oscillation at longer wavelength edge of the attenuation band of the LPFGs at 1111.6 nm. Detail of typical output spectrum is shown in Fig. 4b. The measured FWHM is 0.4 nm and the resolution of the optical spectrum analyzer is 0.1 nm. The observed laser wavelength has long term stability except for pump powers slightly above the threshold.

The dependencies of laser output power versus input pump power are shown in Fig. 5 for several configurations of the laser cavity. It should be noted that the input pump power was controlled by changing laser diode current. In such a case, the pump wavelength changes significantly with the current, at the rate of about 1.37 nm/A. Therefore, the laser curves are not straight lines and their slope increases slowly as the pump wavelength shifts towards peak Yb absorption. The maximum of the slope of 67% is for pump power of 2.5-3.5 W. For higher pump the YDF starts to saturate and correspondingly the slope drops. It can be also inferred from the Fig. 5 that the insertion of LPFGs makes little effect on the laser output power.

Extension of laser oscillation range of Yb-doped fiber lasers towards both shorter and longer wavelengths has attracted recently increased attention [Kurkov07]. Oscillation of our laser at shorter wavelengths is also possible providing that the population inversion is high, e.g., for short YDF [Nilsson04]. However, the pump absorption and correspondingly the power conversion efficiency decrease with YDF shortening. In addition, the effect of strong ASE in YDF significantly lowers the population inversion along the fiber. Therefore, bandstop filter for blocking strong ASE inside the YDF itself would increase the population inversion and allow for higher power conversion efficiency in both ring and Fabry-Perot laser configurations [Peterka09]. This effect would be even more pronounced in case of quasi-distributed filtering when multiple LPFGs would be inscribed along the YDF length. The technique of grating inscription using $CO_2$ laser presented here has an advantage that there is no need for the fiber to be UV photosensitive (e.g., doped by Ge) and thus it can be a suitable way for fabrication of LPFGs directly in the YDF. It would also allow for multiple LPFGs inscription along the YDF fiber length. In preliminary tests of the LPFG inscription in the YDF under test we achieved attenuation bands more than 20 dB deep.

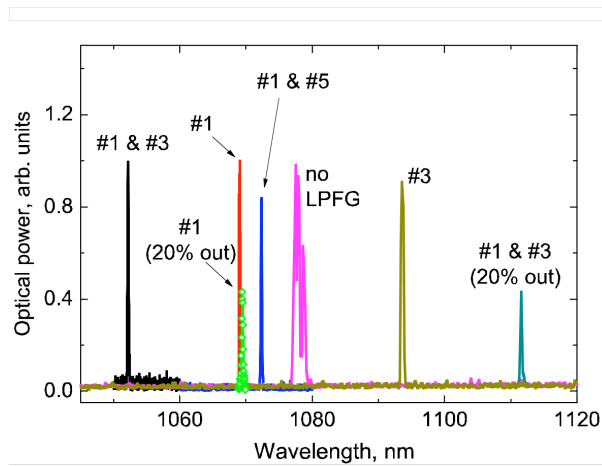 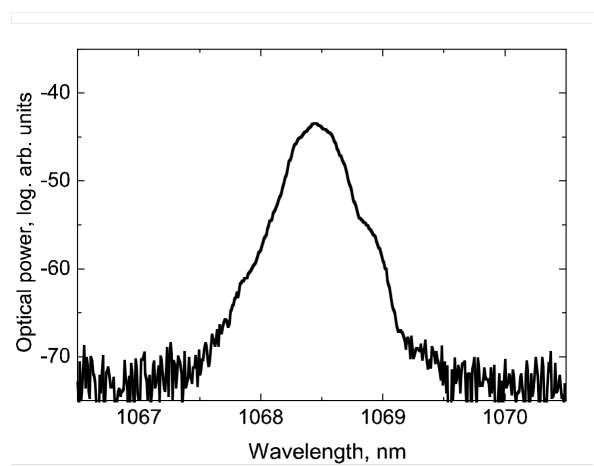

**Figure 4a** Output spectra for various LPFGs combinations. The 80% output coupling was used if not otherwise stated.

**Figure 4b** Detail of typical output spectrum (LPFG #1, 80% out).

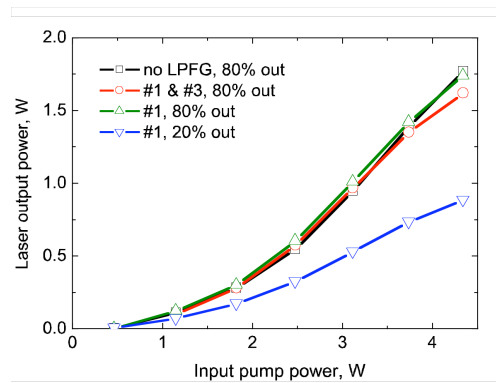

**Figure 5** Dependences of the Yb-doped fiber laser output power versus pump power for several cavity configurations.

## 4. Conclusion

We have demonstrated for the first time selection of operating wavelength of the Yb-doped fiber-ring laser using LPFGs. The LPFGs were inscribed by $CO_2$ laser that allows inherent high power handling of the fabricated spectral filters. The grating period down to 175 μm was achieved being, to our best knowledge, the shortest reported LPFG period using this technique. The spectral range of oscillation between 1050-1110 nm was shown experimentally. This spectral range can be extended to both shorter and longer wavelengths with proper design of the LPFG and length of the laser cavity. In case of laser emission beyond the conventional Yb-doped fiber laser spectra, insertion of LPFGs is also important as they block the ASE around the peak gain and improve the power conversion efficiency.

*Acknowledgments* This work was supported by the French-Czech program Barrande (MEB 020830) and by the Czech Science Foundation (GA102/07/0999) and partially also by the Czech Ministry of Education, Youth and Sports (OC09059 and MSM6840770022) and European Action COST 299.